\documentclass{iaus}
\usepackage{graphicx}

\title[Thermosphere and exosphere of Hot-Jupiters] 
{Thermosphere and exosphere of Hot-Jupiters}

\author[Alain Lecavelier des Etangs]   
{Alain Lecavelier des Etangs$^1$
}

\affiliation{$^1$Institut d'astrophysique de Paris, CNRS/UPMC, 98bis bld Arago, F-75014 Paris, France
}

\pubyear{2008}
\volume{253} 
\pagerange{101--111}
\jname{Transiting Planets}
\editors{F. Pont, ed.}
\begin{document}

\maketitle

\begin{abstract}
Here we describe the observations and the resulting
constraints on the upper atmosphere (thermosphere and exosphere) of the ''Hot-Jupiters''. In particular,  
observations and theoretical modeling of Hot-Jupiter evaporation
are described. 
The observations allowed the discovery that the planet
orbiting HD209458 has an extended atmosphere of escaping
hydrogen and showed the presence of oxygen and carbon
at very high altitude. These observations give unique
constraints on the escape rate and mechanism in the
atmosphere of these planets.
The most recent Lyman-alpha HST observations of 
HD189733b allows for the
first time to compare the evaporation from two different
planets in different environments.
Models to quantify the escape rate from the measured
occultation depths, and an energy diagram to describe
the evaporation state of Hot-Jupiters are presented.
Using this diagram, it is shown that few already known
planets could be remnants of formerly giant planets.
\end{abstract}

\firstsection 

\section{Thermosphere of HD\,209458\,b and HD\,189733\,b}

Physical parameters of the upper atmospheres up to the exosphere (or the thermosphere) can be determined using absorption spectroscopy of transits.
This technique has been developed in details for HD\,209458\,b (Sing et al. 2008a, 2008b; Lecavelier des Etangs et al. 2008b; D\'esert et al.\ 2008) where
the detailed Temperature-Pressure-altitude profile has been estimated from $\sim$0.1\,mbar to $\sim$50\,mbar. In particular, because the atmospheric scale height is directly related to the temperature, the temperature can be easily determined by measurement of variation of transit occultation depth as a function of wavelength. For instance, when detected the slope of the Rayleigh scattering allows a direct
determination of the temperature at the altitude where Rayleigh scattering is optically thick (Lecavelier des Etangs et al.\ 2008a, 2008b).
For more details see the Sing et al.\ contribution in this book (Sing et al. 2008c). 

\section{HD\,209458\,b, an evaporating planet}
\label{HD209458}

Transit observations have also revealed evaporation of the hot-Jupiters closed
to their parent stars. For more than ten years, transit observations allowed discoveries, 
detection, and characterization of extrasolar objects (Lecavelier des Etangs et al.\ 1995, 
1997, 1999a, 1999b, 2005; Lamers et al.\ 1997; Nitschelm et al.\ 2000; 
H\'ebrard \& Lecavelier des Etangs 2006; Ehrenreich et al. 2006, 2007). 
In the recent discoveries, the evaporation of Hot-Jupiters opens a new field of research in 
the exoplanet field. The field was open with the observational
discovery that HD20958b is evaporating (Vidal-Madjar et al.\ 2003, hereafter VM03). This discovery has been challenged by a recent 
work of BenJaffel (2007); but the apparent discrepancy has been solved and the result obtained from this observation data set is strengthened (Vidal-Madjar et al.\ 2008).

Three transits of HD209458b were surveyed with the STIS
spectrograph on-board HST ($\sim$20~km.s$^{-1}$ resolution). 
For each transit, three consecutive HST
orbits were scheduled such that the first orbit ended before the
first contact to serve as a reference, the two following ones
being partly or entirely within the transit.
An average 15$\pm$4\% (1$\sigma$) relative intensity drop
near the center of the Lyman-$\alpha$ line was observed during the transits.
This is larger than expected for the atmosphere of a planet
occulting only $\sim$1.5\% of the star. 

Because of the small distance (8.5~R$_*$) between the planet and the star
(allowing an intense heating of the planet and its classification
as a ``hot Jupiter'') the Roche lobe is only at 2.7 planetary radii 
(i.e. 3.6~R$_{\rm Jup}$ ). Filling up this lobe with hydrogen atoms gives a
maximum absorption of $\sim$10\% during planetary transits. Since
a more important absorption was detected, hydrogen atoms cover a
larger area corresponding to a spherical object of 4.3~R$_{\rm Jup}$ .
Observed beyond the Roche limit, these hydrogen atoms must be escaping 
the planet. Independently, the spectral absorption width, with 
blue-shifted absorption up to -130\,km.s$^{-1}$ also shows that some
hydrogen atoms have large velocities relative to the planet, exceeding
the escape velocity. This further confirms that hydrogen atoms 
must be escaping the planetary atmosphere.

\begin{figure}[!tbh]
\begin{center}
\includegraphics[height=\textwidth,angle=-90]{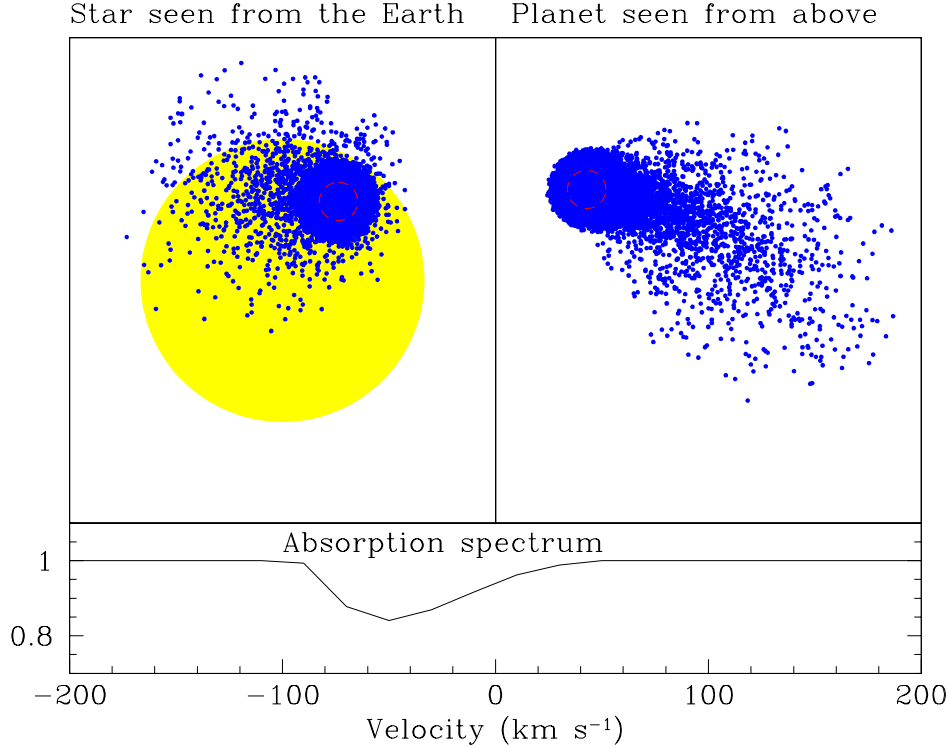}
\caption{A numerical simulation of
hydrogen atoms sensitive to radiation pressure 
(0.7 times the stellar gravitation) above an altitude
of 0.5 times the Roche radius where the density is assumed to be
2$\times$10$^{5}$~cm$^{-3}$ is presented here. It corresponds to
an escape flux of $\sim 10^{10}$~g~s$^{-1}$. The mean
ionization lifetime of escaping hydrogen atoms is 4 hours. The model
yields an atom population in a curved cometary like tail 
(see details in Vidal-Madjar \& Lecavelier 2004).}
\end{center}
\label{fig1}
\end{figure}

The observed 15\% intensity drop could only be explained if
hydrogen atoms are able to reach the Roche lobe of the planet and
then escape. To evaluate the amount of escaping atoms
a particle simulation was built, in
which hydrogen atoms are assumed to be sensitive to the
stellar gravity and radiation pressure (see Fig.~\ref{fig1}). In this
simulation, escaping hydrogen atoms expand in an asymmetric
cometary like tail and are progressively ionized when moving away
from the planet. Atoms in the evaporating coma and tail cover a
large area of the star.
An escape flux of $\sim$10$^{10}$~g.s$^{-1}$ is needed to explain 
the observations. Taking into account the tidal forces and 
the temperature rise expected in the upper atmosphere, 
theoretical evaluations are in good agreement with the observed rate
(see references in Sect~\ref{diagram}).

\section{Hydrodynamical escape or ``Blow-off''}
\label{Blow-off}

Four transits of HD\,209458\,b were then observed,
again with the STIS spectrograph on board HST, but at lower resolution
(Vidal-Madjar et al. 2004, hereafter VM04). 
The wavelength domain
(1180-1710\AA ) includes H\,{\sc i} as well as C\,{\sc i} , C\,{\sc i} , C\,{\sc i} , 
N\,{\sc v} , O\,{\sc i} , S\,{\sc i}, Si\,{\sc iii}, 
S\,{\sc iv} and Fe\,{\sc ii} lines. During the transits, absorptions are
detected in H\,{\sc i} , O\,{\sc i} and C\,{\sc ii} (5$\pm$2\%, 10$\pm$3.5\% and 6$\pm$3\%,
respectively). No absorptions are detected for other lines. The 5\% mean
absorption over the whole H\,{\sc i} Lyman-alpha line is consistent with the previous
detection at higher resolution (VM03). The absorption
depths in O\,{\sc i} and C\,{\sc ii} show that
oxygen and carbon are present in the extended upper atmosphere of
HD\,209458b. These species must be carried out up to the Roche lobe and
beyond, most likely in a state of hydrodynamic escape.

\section{A diagram for the evaporation status of extrasolar planets}
\label{diagram}

The observational constraints given in previous sections 
have been used to developed a large
number of models. These models aim at a better understanding of 
the observed escape rate and evaporation properties, 
and subsequently drawing the consequence on
other planets and planetary systems (Lammer et al. 2003; 
Lecavelier des Etangs et al. 2004, 2007; Baraffe et al. 2004, 2005, 2006; 
Yelle 2004, 2006; Jaritz et al. 2004; Tian et al. 2005; Hubbard et al. 2005; 
Garcia-Munoz 2007). However, all the modeling efforts lead to the
conclusion that most of the EUV and X-ray input energy by
the harboring star is used by the atmosphere to escape the planet 
gravitational potential.

\begin{figure}[!htbp]
\begin{center}
\includegraphics[height=\textwidth,angle=90]{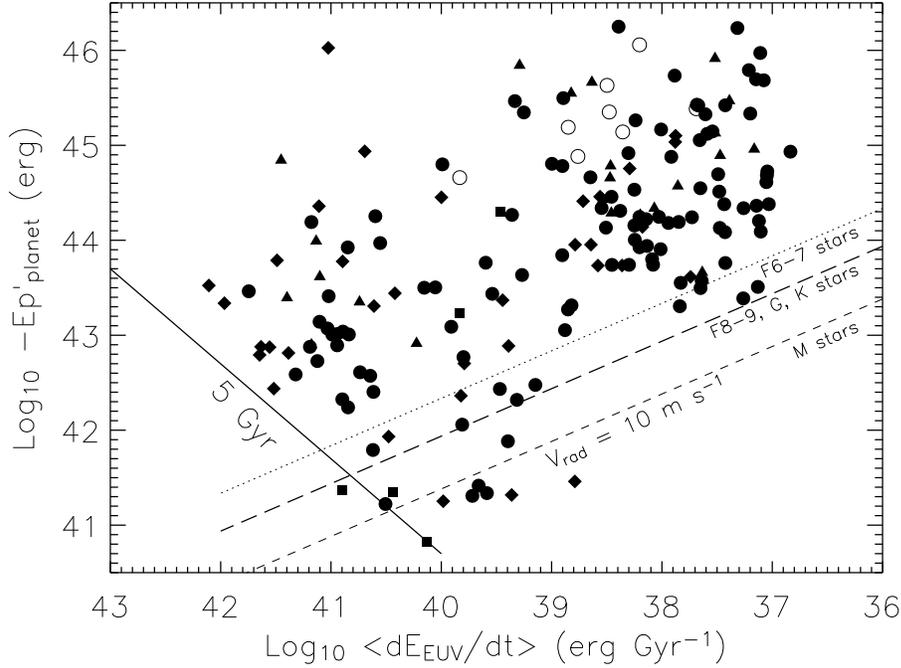}
\caption[]{
Plot of the potential energy of the extrasolar planets as a function of the 
mean EUV energy received per billion of years, $<dE_{\rm EUV}/dt>$.
To keep the planets with the smallest orbital distances 
in the left part of the diagram, the direction of the abscissa axis is chosen 
with the largest value of the mean energy flux toward the left.
Identified planets are plotted with symbols depending on the type of the
central star: triangles for F stars, filled circles for G stars, diamonds for K stars
and squares for M stars; planets orbiting class III stars are plotted with empty circles.     
From the position in the diagram, the typical lifetime of a given planet can be 
rapidly extracted. If the mean energy flux $<$$dE_{\rm EUV}/dt$$>$
is given in unit of {\it erg per billion years}, 
and the potential energy is given in unit of {\it erg}, 
the simple ratio of both quantities provides the
corresponding lifetime in billion of years. 
In the diagram, lifetime isochrones are straight lines.
The lifetime of 5\,Gyr is plotted with a thick line. 

The striking result is the absence of planets in the bottom left region 
which corresponds to light planets (small $-E'_p$) at short orbital distances 
(large $<dE_{\rm EUV}/dt>$).
A plot of the lifetime line at $t$=5\,Gyr, shows that there are no planets in
this part of the diagram simply because this is an evaporation-forbidden region.
Planets in this region would receive more EUV energy than needed to fill the
potential well of the planet, and evaporate in less than 5\,Gyr, leaving a remaining core,
an evaporation remnant (also named a ``chthonian'' planet; Lecavelier des Etangs 
et al.\ 2004).
}
\label{fig2}
\end{center}
\end{figure}

Therefore, to describe the evaporation status of the extrasolar planets, 
an energy diagram as been developed 
in which the potential energy of the planets is plotted versus the
energy received by the upper atmosphere (Lecavelier des Etangs 2007).
This description allows a quick estimate of both the escape rate of the 
atmospheric gas and the lifetime of a planet against the evaporation process. 
In the energy diagram, there is an evaporation-forbidden region in which a gaseous planet 
would evaporate in less than 5 billion years. With their observed characteristics, 
all extrasolar planets are found outside this evaporation-forbidden region
(Fig.~\ref{fig2}).

\section{Neptune mass planets in the diagram}
\label{Neptune mass planets in the diagram}

A plot of the mass distribution of the extrasolar planets shows that Neptune-mass and Earth-mass planets play
a particular role (Lecavelier des Etangs 2007). 
On June 25, with radial velocity searches, nineteen (19!) planets have been found  with mass below 0.072\,$M_{\rm Jup}$ (23~Earth-mass), 
while only one (1!) planet has been identified with mass in the range 0.072\,$M_{\rm jup}$-0.11\,$M_{\rm jup}$ (23-35~Earth-mass). 
This gap is not a bias in the radial velocity searches, 
since more massive planets are easier to detect. 
This reveals the different nature of these
Neptune mass planets orbiting at short orbital distances. But their nature is still a matter of 
debate (Baraffe et al.\ 2005). In particular the question arises 
if they can be the remnants of evaporated more massive
planets (``chthonian planets'') as foreseen in Lecavelier des Etangs et al.\ (2004) ?
Other possibilities include gaseous Neptune-like planets, super-Earth (Santos et al.\ 2004) 
or ocean-planets (Kutchner 2003, L\'eger et al.\ 2004).

\begin{figure}[!htbp]
\begin{center}
\includegraphics[height=\textwidth,angle=90]{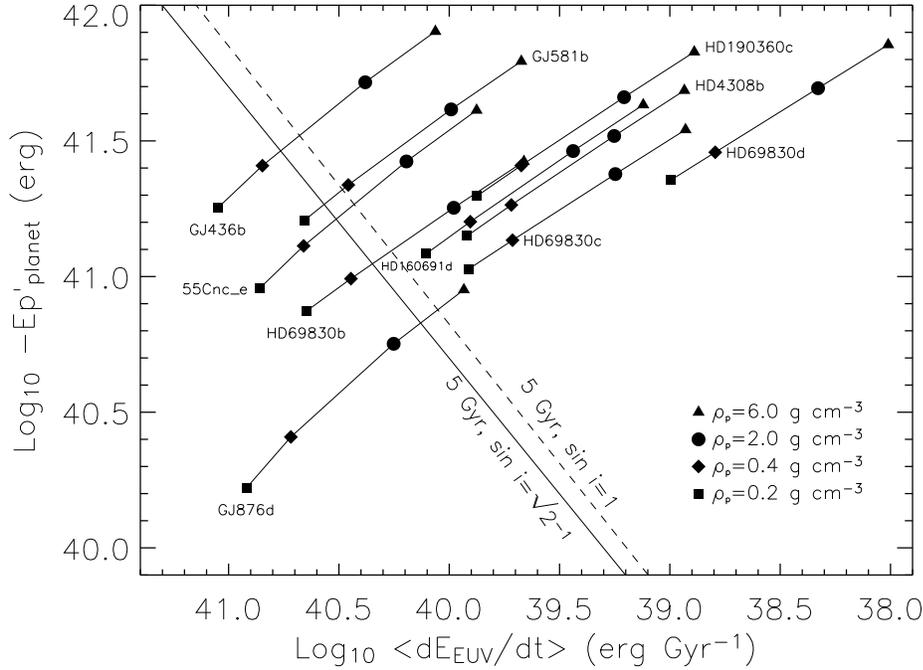}
\caption{Plot of the potential energy of the Neptune mass planet 
as a function of the EUV flux for various planets' density. 
For GJ\,581\,b, GJ\,436\,b, HD\,69830\,b, 55\,Cnc\,e and GJ\,876\,d, and assuming 
$\sin i=\sqrt{2^{-1}}$, lifetime shorter than 5\,Gyr are obtained 
for densities below 0.28, 0.55, 0.56, 0.69 and 3.1\,g\,cm$^{-3}$, respectively.
If $\sin i$=1 (dotted line), the critical (minimum) densities are 
increased to 0.38, 0.74, 0.78, 0.93 and 4.2\,g\,cm$^{-3}$, 
for GJ\,581\,b, GJ\,436\,b, HD\,69830\,b, 55\,Cnc\,e, and GJ\,876\, respectively.
\label{fig3}}
\end{center}
\end{figure}

We plotted the position of these Neptune mass planets in the energy diagram 
with different hypothesis on their density (Fig.~\ref{fig3}). 
We used mean planetary density of $\rho_p$=6\,g\,cm$^{-3}$ for a typical density of 
refractory-rich planets which should describe the chthonian and super-Earth planets. A lower density 
on the order of $\rho_p$=2\,g\,cm$^{-3}$ can be considered as more plausible for volatile-rich
planets describing the ocean planets. For gas-rich planets we assumed much lower density
of $\rho_p$=0.2\,g\,cm$^{-3}$ and $\rho_p$=0.4\,g\,cm$^{-3}$, describing planets which should
look more like irradiated Neptune-like planets.

GJ\,876\,d cannot have the density of an ocean planet and 
needs to be dense enough to be located
above the $t=5$\,Gyr lifetime limit. This planet requires
a density larger than 3.1\,g\,cm$^{-3}$ for its atmosphere to survive. 
GJ\,876\,d could be a big rocky planet, like a super-Earth, or a refractory remnant of
a previous more massive planet (an evaporated ocean planet?).

In brief, the energy diagram allows us to trace three different categories for
the presently identified Neptune mass planets. For half of them, 
the EUV input energy seems not strong enough to affect significantly these 
planets; we cannot conclude on their nature. 
For at least three other planets (GJ\,436\,b, 55\,Cnc\,e and 
HD\,69830\,b), it appears that
they cannot be a kind of low mass gaseous planets. With density necessarily 
above 0.5\,g\,cm$^{-3}$ to survive evaporation,
these planets must contain 
a large fraction of solid/liquid material. Finally, GJ\,876\,d must be 
dense enough, with a density larger than $\sim$3\,g\,cm$^{-3}$, 
to survive the strong EUV energy flux from its nearby parent 
star. This planet must contain a large fraction of massive elements (Fig.~\ref{fig3}).

\section{Observations of HD\,189733\,b}

The observation of the HD209458b transits revealed that the
atmosphere of this planet is hydro-dynamically escaping (Sect.~\ref{HD209458} and~\ref{Blow-off}). 
These observations raised the question of the
evaporation state of hot-Jupiters. Is the evaporation specific to HD\,209458\,b or general to hot-Jupiters? What is
the evaporation mechanism, and how does the escape rate depend on the planetary system characteristics?
The recent discovery of HD\,189733\,b, a planet transiting a bright and nearby K0 star (V=7.7), offers the
unprecedented opportunity to answer these questions. Indeed, among the stars harboring transiting planets,
HD189733 presents the largest apparent brightness in Lyman-$\alpha$, providing capabilities to constrain the
escape rate to high accuracy.

An HST program has been developed to observed HI, CII and OI stellar emission lines to search for atmospheric
absorptions during the transits of HD\,189733\,b. A preliminary analysis of this program has been presented during the conference.
The preliminary H\,{\sc i} Lyman-$\alpha$ 
transit light curve constrains the escape rate in the few 10$^8$\,g\,s$^{-1}$ domain. 
But a more detailed analysis, in particular of the complete data set to cover the whole light curve is needed
to obtain firm conclusion.

In short, the atmosphere of only a single extrasolar planet has been detected so far leading to the discovery of
the evaporation of a hot-Jupiter. Observations of other cases under various physical conditions provide
important constraints on the evaporation state and mechanisms. HD189733b being a very short period planet
orbiting a nearby late type star with bright chromospheric emission lines, it is by far the best target to make
significant progress in that field. Results are awaiting for the coming months.

\section{Conclusion}

In summary, the observation of HI Lyman-$\alpha$ transit 
allows the detection of escaping atmosphere of HD\,209458\,b.
The escape rate has been estimated through modeling 
of the observed transit light curve, given as estimate around 10$^{10}$\,g\,s$^{-1}$.
The detection of heavy elements has then constrained 
the escape mechanism to be an hydrodynamical escape, or ``blow-off''.
Recently, the detection of Balmer jump
has given a new constraint on the temperature of the upper atmosphere
(Ballester et al. 2007).
Finally, an energy diagram allows putting constraints on 
the density of few Hot-Neptunes 
(some of which may be evaporation remnant, or ``chthonian planets'').
The most recent HST observations show that HD18973\,b 
(a very-Hot Jupiter orbiting a K star)
very likely also 
presents a significant evaporation, analysis of these observations 
are still in progress.

\end{document}